\documentclass[10pt,superscriptaddress,twocolumn,amsmath,amssymb,aps,pra,showpacs]{revtex4-1}
\usepackage{mathrsfs}
\usepackage{graphicx}
\usepackage{dcolumn}
\usepackage{bm}
\usepackage[title,titletoc,header]{appendix}

\usepackage{amssymb}
\usepackage{amsmath}
\usepackage{mathrsfs}
\usepackage{amsmath}
\usepackage{subfigure}
\usepackage{float}
\usepackage[title,titletoc,header]{appendix}
\usepackage{graphicx}

\newcommand{\be}{\begin{equation}}
\newcommand{\ee}{\end{equation}}
\newcommand{\bea}{\begin{eqnarray}}
\newcommand{\eea}{\end{eqnarray}}

\begin{document}
\title{Exact mobility line and mobility ring in the complex energy plane of a flat band lattice with a  non-Hermitian quasiperiodic potential}
\author{Guang-Xin Pang}
\affiliation{School of Physics and Materials Science, Guangzhou University, Guangzhou 510006, China}

%
%


\author{Zhi Li}
\affiliation{Key Laboratory of Atomic and Subatomic Structure and Quantum Control (Ministry of Education),
Guangdong Basic Research Center of Excellence for Structure and Fundamental Interactions of Matter,
School of Physics, South China Normal University, Guangzhou 510006, China}
\affiliation{Guangdong Provincial Key Laboratory of Quantum Engineering and Quantum Materials,
Guangdong-Hong Kong Joint Laboratory of Quantum Matter,
Frontier Research Institute for Physics, South China Normal University, Guangzhou 510006, China}

\author{Shan-Zhong Li}
\affiliation{Key Laboratory of Atomic and Subatomic Structure and Quantum Control (Ministry of Education),
Guangdong Basic Research Center of Excellence for Structure and Fundamental Interactions of Matter,
School of Physics, South China Normal University, Guangzhou 510006, China}
\affiliation{Guangdong Provincial Key Laboratory of Quantum Engineering and Quantum Materials,
Guangdong-Hong Kong Joint Laboratory of Quantum Matter,
Frontier Research Institute for Physics, South China Normal University, Guangzhou 510006, China}
\author{Yan-Yang Zhang}
\affiliation{School of Physics and Materials Science, Guangzhou University, Guangzhou 510006, China}
\author{Jun-Feng Liu}
\affiliation{School of Physics and Materials Science, Guangzhou University, Guangzhou 510006, China}

\author{Yi-Cai Zhang}\thanks{Contact author: E-mail:zhangyicai123456@163.com}
\affiliation{School of Physics and Materials Science, Guangzhou University, Guangzhou 510006, China}


\date{\today}
\begin{abstract}
%
   In this study, we investigate the problem of Anderson localization in a one-dimensional flat band lattice with a non-Hermitian quasiperiodic on-site potential.  First of all, we discuss the influences of non-Hermitian potentials on the existence of critical states. Our findings show that, unlike in Hermitian cases, the non-Hermiticity of the potential leads to the disappearance of critical states and critical regions. Furthermore, we are able to accurately determine the Lyapunov exponents and the mobility edges. Our results reveal that the mobility edges form mobility lines and mobility rings in the complex energy plane. Within the mobility rings, the eigenstates are extended, while the localized states are located outside the mobility rings. For mobility line cases, only when the eigenenergies lie on the mobility lines, their corresponding eigenstates are extended states.
   Finally, as the energy approaches the mobility edges, we observe that, differently from Hermitian cases, here the critical index of the localization length is not a constant, but rather varies depending on the positions of the mobility edges.


%


\end{abstract}

\maketitle
\section{Introduction}
 It is believed that even a weakly uncorrelated  disorder in one and two dimensions \cite{Economou2006} can lead to Anderson localization \cite{Anderson1957}. In three dimensional disorder system, there is a mobility edge $E_c$ that separates localized states from delocalized states \cite{Economou1972}.
  Near localized-delocalized transition points ($E_c$),  the localization length $\xi(E)$ diverges, i.e.,
\begin{align}\label{10}
\xi(E)\propto|E-E_c|^{-\nu}\rightarrow\infty, \ \ as \ E\rightarrow E_c,
\end{align}
where  $\nu$ is the critical index \cite{Huckestein1990}.
 Interestingly, if the disorder is correlated, the one-dimensional system can have extended states \cite{Zhangwei2004} and undergo a localized-extended transition \cite{Cheraghchi2005}.
Furthermore, the correlation between disorders  can even lead to the emergence of mobility edges in one-dimensional lattice models \cite{Flores}.


Later on, the investigations of the localized-delocalized transition and mobility edge have been extended to  the cases of quasiperiodic potential \cite{42,Thouless,Kohmoto1983,Kohmoto2008,Cai2013,Roati2008,Lahini2009,Deng2019,Cao,Goncalves1,Vu,Goncalves2,xu2020,lli2020,an,zwang,Xiaoshui}.
A famous example where the localized-extended transition can occur is the Aubry-Andr\'{e} lattice model (AA model) \cite{42}.
However, this  system does not have mobility edges.
The nonexistence of mobility edges originates from the exact self-duality of this
model at the critical point. In general, the breaking of
the self-duality would result in the appearance of mobility
edges in the one dimension system \cite{32,33,34,35,43,44,45,46,47,48}.

Ganeshan, Pixley and Das Sarma proposed a generalized Aubry-Andr\'{e} model (GAA model) which
has mobility edges \cite{36,49}. The GAA model is
 \begin{align}\label{1}
&t[\psi(n+2)+\psi(n-2)]+\frac{2\lambda \cos(2\pi\beta n+\phi)}{1-\alpha \cos(2\pi\beta n+\phi)}\psi(n)\notag\\
&=E\psi(n).
\end{align}
where $t$ is hopping, $n\in Z$  is lattice site index, $\lambda$ describes
the quasi-periodic potential strength, $\beta$ is an irrational
number, $\phi$ is a phase, $\alpha$ is a real number.
Importantly, its mobility edges can be exactly determined by a generalized self-duality transformation \cite{49}.
 A mosaic lattice model with quasiperiodic disorder has been proposed \cite{Wangyucheng2020}. This model has been found to have mobility edges, which can be exactly determined using Avila's theory \cite{Liu2021, Avila2015,Zhou2023}.

When $|\alpha|<1$, the quasi-periodic potential is bounded [see Eq.(\ref{1})], and the system can undergo a localized-extended transition. However, when $|\alpha|\geq1$, the potential is unbounded, resulting in an unbounded energy spectrum. In this case, the system exhibits critical states and undergoes localized-critical transitions \cite{yicai}. Additionally, there is a critical region in the parameter plane that consists of critical states. As the energy approaches the localized-critical transition point (named as the anomalous mobility edge \cite{Liu2022}), the critical index of the localized length is $\nu=1/2$, which is different from the $\nu=1$ observed in the bounded case ($|\alpha|<1$) \cite{yicai}.

 Recently, a flat band lattice model was proposed  \cite{zhangyicai2024}, which has much richer physics compared to the original Ganeshan-Pixley-Das Sarma's GAA model.
 It was found that the physics of localization for both the bounded and unbounded quasi-periodic potentials can be realized in this flat band lattice model.
  Not only can localized states, extended states, and critical states coexist, but the system also exhibits multiple transitions between these states, such as localized-extended, localized-critical, extended-critical, and localized-localized transitions.

The studies of Anderson localization in quasi periodic potentials have been extended to non-Hermitian situations \cite{Jazaeri2001,Jiang2019,lliu2021,lliu2020,Schiffer,cai2021,LiuB2021,Zeng2020,Yuce,Ccai2021,Longhi2021,zxu,han2020,tliu,chen2020,lzhou,xxia,Gandhi,Padhan,qliu,lzhang,Acharya,Vatnik2017,xiangping}.
  In a non-Hermitian quasiperiodic system, the concept of  mobility edge has been generalized to the complex energy plane \cite{Liu2020,Lizhi2024, Wangli2024,WangL,shanzhong24}, named as a mobility ring \cite{Lizhi2024}. However,  exactly analytical results on complex mobility edges are relatively scarce, particularly in non-Hermitian quasiperiodic flat bands. Therefore, it is necessary to extend the study of mobility edges in flat bands to the complex plane in order to explore their physical properties.

  One may wonder what would happen if we generalize the above flat model to the non-Hermitian case. What would be the fate of critical states and  critical regions under the influences of non-Hermitian potentials? Are there any mobility rings?

 In this study,
we aim to address the aforementioned questions by expanding upon the flat model  to include non-Hermitian quasperiodic potentals. Our findings reveal that the presence of a non-Hermitian on-site potential results in the disappearances  of critical states and critical regions. We provide a thorough explanation for this observation and its underlying mechanism. In such scenarios, the system only exhibits extended and localized states. Compared to localized states, the spatial extensions of  extended states are significantly larger.

Additionally, we accurately determine the Lyapunov exponent $\gamma(E)$ using Avila's theory. This allows us to precisely map out the phase diagram in the complex energy plane and identify mobility edges. Interestingly, we observe that these mobility edges can form not only ring structures, but also line segment structures, which we refer to as mobility lines. Finally, we note that the critical index of localized length in the complex energy plane is dependent on the positions of the mobility edges.

\begin{figure}
\begin{center}
\includegraphics[width=0.8\columnwidth]{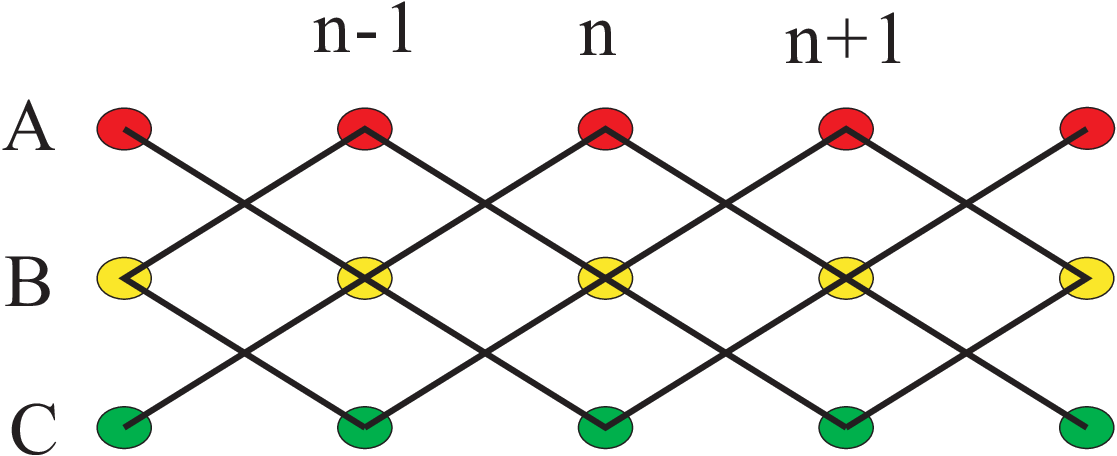}
\end{center}
\caption{ The figure shows a one-dimensional flat band lattice, composed of two intersecting diamond lattices. The solid lines represent the hopping between lattice sites. The sublattice indices are denoted by $A, B, C$, and the unit cell indices are represented by $n-1, n, n+1$.}
\end{figure}

The work is organized as follows. First, the flat band  model and its three energy bands are given in Sec.\textbf{II}.  Next, we  investigate the localization problem of quasi-periodic potential of ``type III" and ``type II" in Sec.\textbf{III} and Sec.\textbf{IV}.
 Finally, a summary is given in Sec.\textbf{V}.


\section{ The flat band model}
In this work, like reference \cite{zhangyicai2024}, we consider the following flat band lattice model consisting of three sublattices A, B, and C, i.e.,
 \begin{align}
&H=H_0+V\notag\\
&H_0=-\frac{it}{\sqrt{2}}\sum_{n\in Z}[a^{\dag}_{n-1}b_{n}+b^{\dag}_{n-1}a_{n}+b^{\dag}_{n-1}c_{n}+c^{\dag}_{n-1}b_{n}]+\emph{h.c.}\notag\\
&+m\sum_{n\in Z}[a^{\dag}_{n}a_{n}-c^{\dag}_{n}c_{n}],
\end{align}
where $V_p$ is potential energy, integer $n$ is the unit cell's index, $H_0$ is the free-particle Hamiltonian, $t>0$ is hopping parameter, and $m>0$ is energy gap parameter.
$a(b/c)_{n}$ are the annihilation operators for the states in sublattices $A(B/C)$, respectively.
The flat band lattice structure is depicted in Fig. 1. It is composed of two intersecting diamond lattices, as shown in the figure.

When potential $V_p=0$, applying a Fourier transform, the free particle Hamiltonian $H_0$ can be written as
   \begin{align}
&H_0=\sqrt{2}t\sum_{-\pi\leq k\leq \pi}\sin(kd)[a^{\dag}_{k}b_{k}+b^{\dag}_{k}a_{k}+b^{\dag}_{k}c_{k}+c^{\dag}_{k}b_{k}]\notag\\
&+m\sum_{-\pi\leq k\leq \pi}[a^{\dag}_{k}a_{k}-c^{\dag}_{k}c_{k}],
\end{align}
where $d$ is lattice constant. In the whole manuscript, we would set $d=1$ for simplifications.

Furthermore, in the above Hamiltonian $H_0$, we can identify the above three sublattices $A,B,C$ as three internal spin components $1,2,3$. After diagonalizing $H_0$,
  we can obtain the three eigenenergies:
\begin{align}
  &E_{-, k}=-\sqrt{4t^2\sin^2(k)+m^2};\notag\\
  &E_{0, k}=0;\notag\\
  &E_{+, k}=\sqrt{4t^2\sin^2(k)+m^2},
\end{align}
where $|-(0/ +), k\rangle$ denote the eigenstates of lower, middle (flat) and upper bands, respectively.
We can see  that a flat band with zero energy ($E_{0, k}=0$) appears between the upper and lower bands.

In the vicinity of momentum $k = 0$, the three-band Hamiltonian described above can be approximated by a continuous spin-1 Dirac model.
The bound state problems of this continuous spin-1 Hamiltonian have been extensively studied with various types of potentials \cite{24,25,29,30,31}. It is found that a short-ranged potential can induce an infinite number of bound states, and the bound state in continuum (BIC) can appear.

\section{Anderson localization in  a non-Hermitian quasiperiodic potential of type III }

In the following, we assume the potential energy $V_p$ has following form in basis $|1,2,3\rangle$, namely,
\begin{align}\label{3}
V_p=\sum_n V_{11}(n)a^{\dag}_{n}a_{n}.
\end{align}
In the whole manuscript, we would refer to such a kind of potential as potential of ``type III" \cite{30}.
The bound state problems with potential of ``type I, II and III" were investigated \cite{24,29,30,31}.

In this work, we take a specific non-Hermitian quasi-periodic potential, namely
\begin{align}
V_{11}(n)=V_0e^{i\theta}\cos(2\pi\beta n+\phi+i\epsilon).
\end{align}
where real number  $V_0$ is potential strength, $\beta$ is a irrational number, $\theta$, $\epsilon$ and $\phi$ are real number. In comparison with Hermitian case \cite{zhangyicai2024}, here the non-Hermiticity enter into system through the overall phase factor $e^{i\theta}$ and imaginy phase $i\epsilon$ in the cos function. In the following,  we take $m=t$, $\beta=\frac{\sqrt{5}-1}{2}$ for the sake of definiteness, and use the units of system of $t=1$.

Using three component wave functions $\psi_1, \psi_2. \psi_3$, the Schr\"{o}dinger equation ($H|\Psi\rangle=E|\Psi\rangle$)  can be expressed by
\begin{align}\label{9}
&\frac{-it}{\sqrt{2}}[\psi_{2}(n+1)-\psi_{2}(n-1)]=[E-m-V_{11}(n)]\psi_{1}(n),\notag\\
&\frac{-it}{\sqrt{2}}[\psi_{1}(n+1)-\psi_{1}(n-1)+\psi_{3}(n+1)-\psi_{3}(n-1)]\notag
\\
&=E\psi_{2}(n),\notag\\
&\frac{-it}{\sqrt{2}}[\psi_{2}(n+1)-\psi_{2}(n-1)]=[E+m]\psi_{3}(n).
\end{align}
Eliminating wave functions of second and third components in Eq.(\ref{9}), we get an effective equation for $\psi_1(n)$, i.e.,
\begin{align}\label{91}
&t^2[\frac{E-V_{11}(n+2)/2}{E+m}\psi_{1}(n+2)-2\frac{E-V_{11}(n)/2}{E+m}\psi_{1}(n)\notag\\
&+\frac{E-V_{11}(n-2)/2}{E+m}\psi_{1}(n-2)]\notag\\
&=-E[E-m-V_{11}(n)]\psi_{1}(n).
\end{align}
Further we introduce an auxiliary
wave function $\psi(n)\equiv\frac{E-V_{11}(n)/2}{E+m}\psi_{1}(n)$,  an effective hopping $\tilde{t}$, an effective total energy $\tilde{E}$, an effective potential strength $\lambda$ and an effective parameter $\alpha$, i.e.,
\begin{align}\label{10}
&\tilde{t}\equiv t^2,\notag\\
&\tilde{E}\equiv-E^2+m^2+2\tilde{t}=-E^2+m^2+2t^2,\notag\\
&\lambda\equiv-\frac{V_0e^{i\theta}(E+m)^2}{4E},\notag\\
&\alpha\equiv\frac{V_0e^{i\theta}}{2E},
\end{align}
 we get an equation for $\psi(n)$
 \begin{align}\label{11}
&\tilde{t}[\psi(n+2)+\psi(n-2)]+\frac{2\lambda \cos(2\pi\beta n+\phi+i\epsilon)}{1-\alpha \cos(2\pi\beta n+\phi+i\epsilon)}\psi(n)\notag\\
&=\tilde{E}\psi(n).
\end{align}
Comparing with Eq.(\ref{1}), it is found that this is an effective generalized Aubry-Andr\'{e} model  whose effective lattice constant is two times of the original lattice constant, i.e., $\tilde{d}=2d=2$.

From Eq. (\ref{10}), we can see that when $\theta=0$ and $E$ is real, the effective parameter $\alpha$ is also real.
Further, when $\epsilon=0$ and the effective parameter $|\alpha|\geq1$, the effective potential $\frac{2\lambda \cos(2\pi\beta n+\phi+i\epsilon)}{1-\alpha \cos(2\pi\beta n+\phi+i\epsilon)}=\frac{2\lambda \cos(2\pi\beta n+\phi)}{1-\alpha \cos(2\pi\beta n+\phi)}$ in Eq.(\ref{11}) becomes unbounded.
In the previous studies  \cite{yicai,zhangyicai2024}, it is found that the unbounded potential would lead to critical states, critical regions, and localized-critical transitions.

However, in the presence of non-Hermitian conditions where $\theta\neq0$ or $\epsilon\neq0$, the critical states and critical regions disappear, leaving only localized and extended states.
This can be explained as follows:
the wave function of an extended state typically spans the entire lattice, while a localized state occupies only a finite number
of lattice sites. The critical state is composed of several disconnected patches that interpolate between the localized and extended states \cite{Liu2022,yicai,zhangyicai2024,yicai2025,shangzhong2025}.
As a result, the wave function of a critical state usually takes a value of zero at the lattice site where it is disconnected.
This is due to the unboundedness of the quasiperiodic potential. When $|\alpha|\geq1$, the denominator of the potential, $|1-\alpha \cos(2\pi\beta n+\phi)|$, can become arbitrarily small at some lattice sites due to the ergodicity of the map $\phi \rightarrow \phi+2 \pi \beta $. This means that for a given eigenenergy $E=E_n$, the potential $\frac{2\lambda \cos(2\pi\beta n+\phi)}{1-\alpha \cos(2\pi\beta n+\phi)}$ can take arbitrarily large values at these sites. As a result, the wave function becomes vanishingly small and disconnected at these lattice sites, leading to the appearance of critical states.

 From these discussions, we can understand that the reason for the occurrence of critical states is that as lattice site index $n$ varies, the denominator of the quasiperiodic potential can approach zero. Roughly speaking, in the case of an unbounded Hermitian potential, the denominator can become zero. However, in the case of non-Hermitian conditions where $\theta\neq0$ or $\epsilon\neq0$, the denominator of the potential $1-\alpha \cos(2\pi\beta n+\phi+i\epsilon)$ becomes a complex number. In order for the denominator to be zero, both the real and imaginary parts must be zero. However, it is highly unlikely for both the real and imaginary parts to be zero at the same time as lattice site index $n$ varies. Therefore, the critical state is unlikely to occur in this case, and the system will only have localized and extended states.

The localized properties of an eigenstate can be characterized by Lyapunov exponent.
 For a given parameter $E$, as $n$ increases, we can assume that the wave function grows exponentially according to a law \cite{Ishii,Furstenberg}, expressed as
\begin{align}
\psi(n)\sim e^{\gamma(E) n}, &\ as \ n\rightarrow \infty,
\end{align}
where $\gamma(E)\geq0$ is Lyapunov exponent which measures the average growth rate of wave function.
If the parameter $E$ is not an eigen-energy of $H$, the Lyapunov exponent will be positive, i.e., $\gamma(E)>0$ \cite{Jonhnson1986}. When $E$ is an eigen-energy of the system, the Lyapunov exponent can be either zero or positive \cite{yicai}, depending on the type of state. Extended and critical states have a Lyapunov exponent of zero, while localized states have a positive exponent.

\begin{figure}
\begin{center}
\includegraphics[width=1.1\columnwidth]{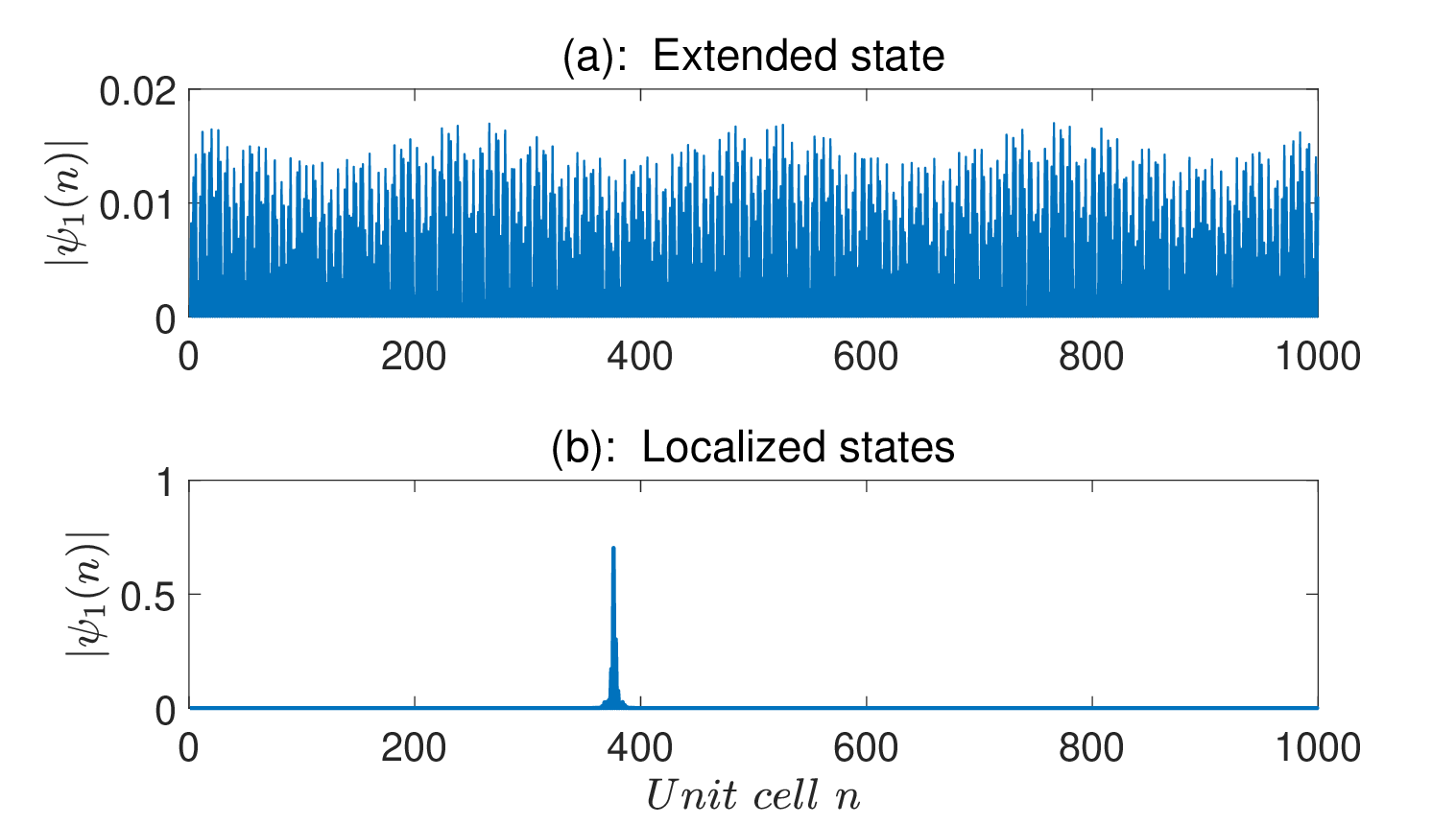}
\end{center}
\caption{Two typical wave functions for extended states and localized states.}
\end{figure}

When $E$ belongs to eigenvalues,  using Avila's theory \cite{Liu2021,Avila2015,51},  we can get the exact expression of Lyapunov exponent , i.e. (for detailed calculation, please refer to the Appendix),
\begin{align}\label{141}
\gamma(E,\epsilon)=
\left\{\begin{array}{cccc}
&&\frac{1}{2}\textrm{Max}\{0,|\epsilon|+\ln(\frac{|P\pm\sqrt{P^2-4\alpha^2}|}{2|1+\sqrt{1-\alpha^2}|})\}, \\
\\
 &&  for \ \ \ \ |\epsilon|<
\ln(|\frac{1+\sqrt{1-\alpha^2}}{\alpha}|)\\
\\
&&\frac{1}{2}\textrm{Max}\{0,\ln(\frac{|P\pm\sqrt{P^2-4\alpha^2}|}{2|\alpha|})\},  \\
\\
&& for  \ \ \ \ |\epsilon| \geq \ln(|\frac{1+\sqrt{1-\alpha^2}}{\alpha}|)\\
  \end{array}\right.
\end{align}
where
\begin{align}
P=\frac{\alpha \tilde{E}+2\lambda}{\tilde{t}}.
\end{align}


To analyze the properties of the eigenstates, we have utilized numerical methods to solve Eq. (\ref{9}). Our approach involves using a total of $N=1000$ unit cells and constructing a $3N\times3N$ matrix with open boundary conditions at the two end sites. This matrix is then diagonalized, resulting in $3\times N=3000$ eigenenergies and eigenstates. Two typical wave functions for extended state and localized state are shown in Fig. (2).

In addition, with Eq.(\ref{11}), the Lyapunov exponent can be calculated numerically with transfer matrix method, i.e.,
\begin{align}\label{V}
&\gamma(E,\epsilon)=\lim_{L \rightarrow \infty }\frac{\ln(|\Psi(2L)|/|\Psi(0)|)}{2L}\notag\\
&=\lim_{L\rightarrow \infty}\frac{\ln(|T(2L)T(2L-2)...T(4)T(2)\Psi(0)|/|\Psi(0)|)}{2L}
\end{align}
where $L$ is a positive integer, transfer matrix
\begin{align}
T(n)\equiv\left[\begin{array}{ccc}
\frac{\tilde{E}}{\tilde{t}}-\frac{2\lambda}{\tilde{t}} \frac{\cos(2\pi\beta n+\phi+i\epsilon)}{1-\alpha \cos(2\pi\beta n+\phi+i\epsilon) } &-1  \\
1&0\\
  \end{array}\right],
\end{align}
\begin{align}
\Psi(n)\equiv\left[\begin{array}{ccc}
\psi(n+2)  \\
\psi(n)\\
  \end{array}\right],
\end{align}
and
\begin{align}
|\Psi(n)|=\sqrt{|\psi(n+2)|^2+|\psi(n)|^2}.
\end{align}

To be specific, after we obtain the corresponding eigenenergies and eigenstates, we then proceed to numerically calculate the Lyapunov exponents for all of these eigenenergies, as shown by  the discrete red  squares in Fig. 3.
 In our numerical calculation, we take $L=200$,  phase $\phi=0$, $\psi(0)=0$ and $\psi(2)=1$ in Eq.(\ref{V}).

Eq. (\ref{141}) has been confirmed by our numerical results, as shown in Fig. 3.
 The blue dots in Fig. 3 are given by Eq. (\ref{141}) for all the eigenvalues.
 It is evident that the two sets of discrete points overlap very well on the graph.


%


\begin{figure}
\begin{center}
\includegraphics[width=1.1\columnwidth]{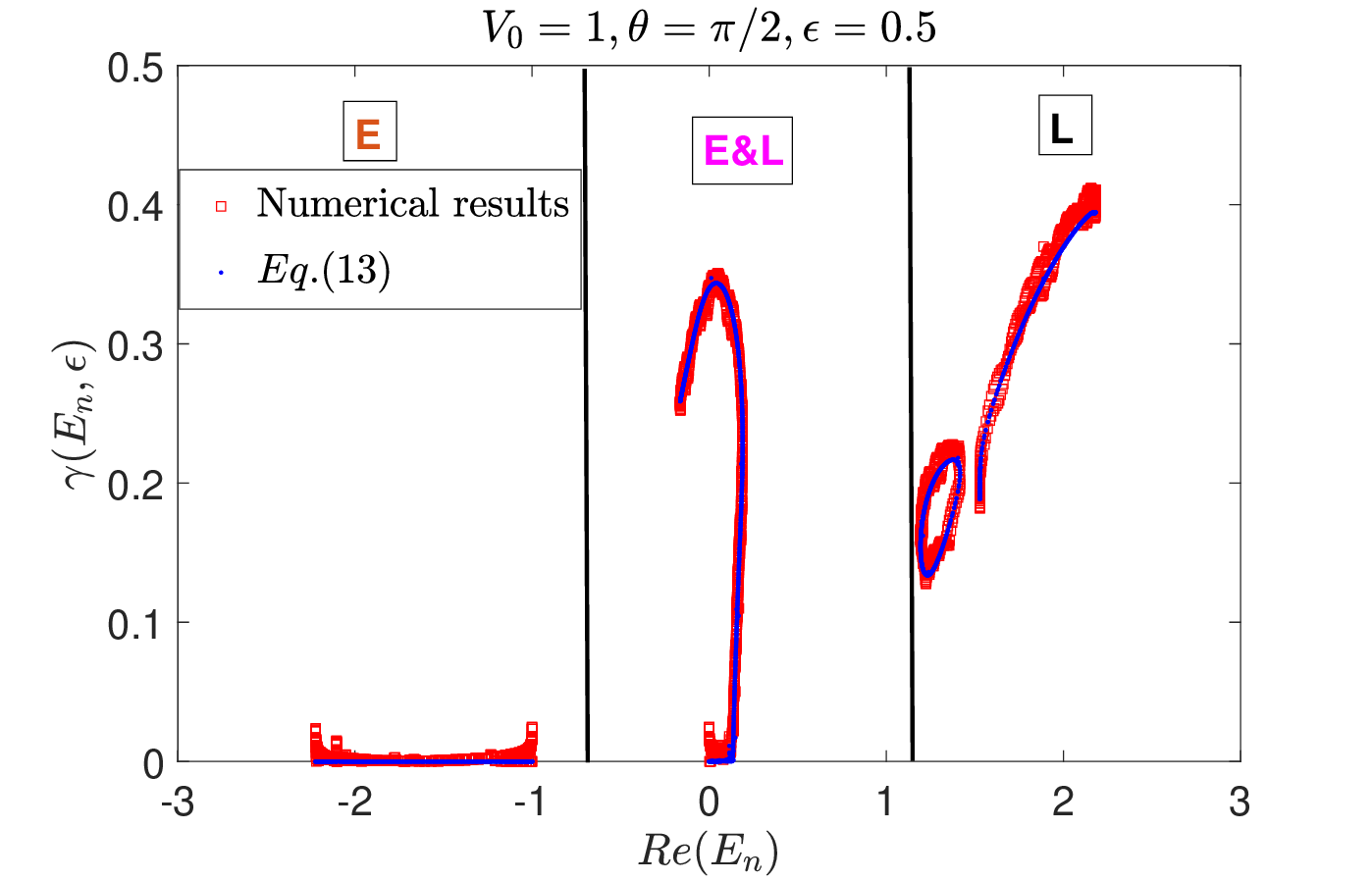}
\end{center}
\caption{Lyapunov exponent for quasi-periodic potential of type III with $V_0=1, \theta=\pi/2, \epsilon=0.5$. The red squares represent the results from numerical calculation using Eq.(\ref{V}), while the blue dots are obtained from Eq.(\ref{141}). It is evident that the two sets of discrete points overlap very well. Here $\textbf{L}$, $\textbf{E}$ and $\textbf{E}$ \& $\textbf{L}$ denote localized states, extended states and the mixture of localized and extended states, respectively. It is evident that the Lyapunov exponents of localized states are positive, while the Lyapunov exponents for extended states are zero (also see Figs. 4 and 5). }
\end{figure}

\begin{figure}
\begin{center}
\includegraphics[width=1.1\columnwidth]{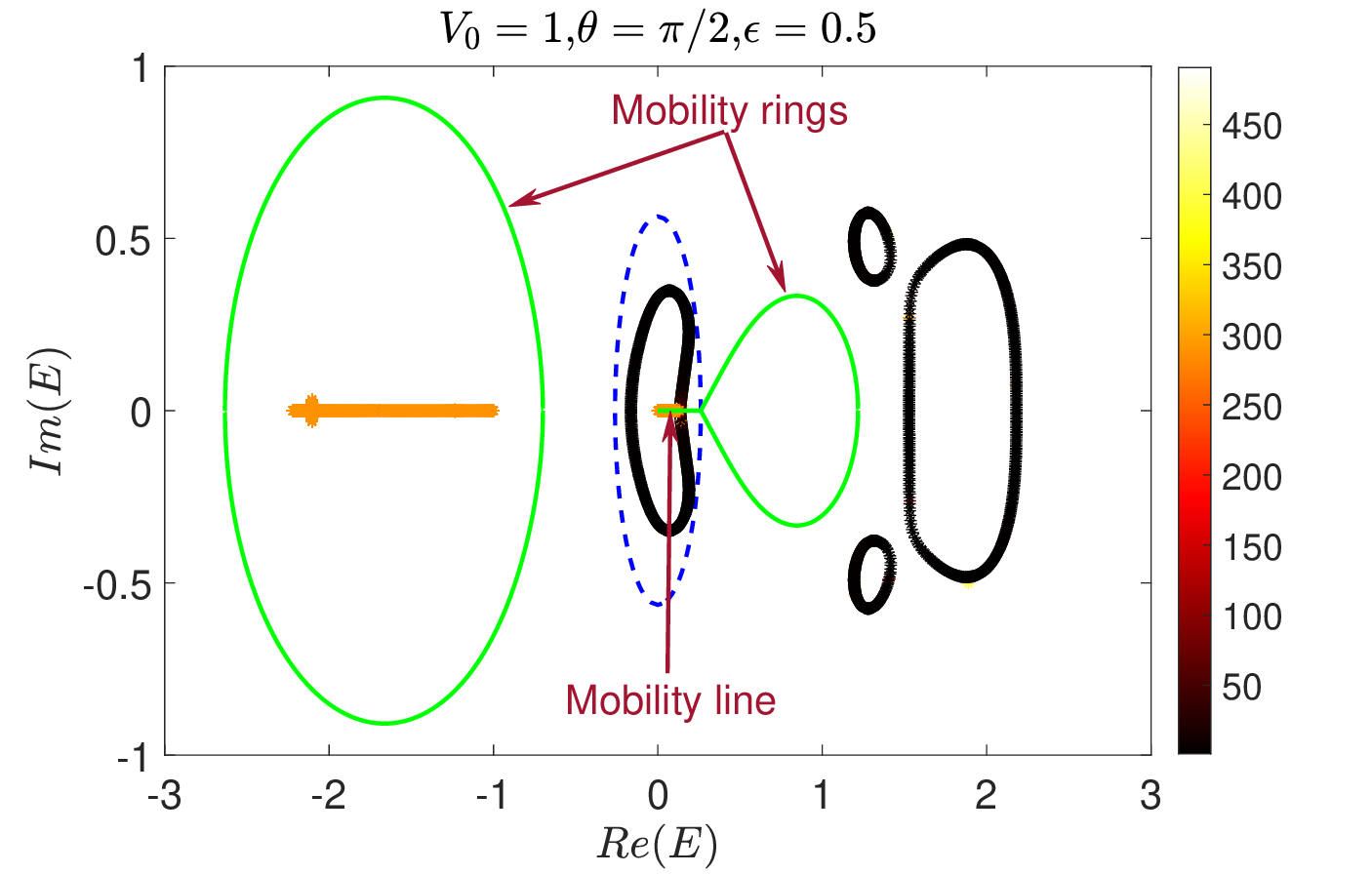}
\end{center}
\caption{ The phase diagram for quasi-periodic potential of type III in the complex number $E$ plane for $V_0=1$, $\theta=\pi/2$, $\epsilon=0.5$ and lattice size $3\times N=3\times 1000=3000$.
  The phase boundaries, also known as mobility edges $E_c$, are indicated by the green solid line  segment and curves, and their equations are given by Eqs. (\ref{24}) and (\ref{30}). The states within the green loops and on the green line segment (yellow points)  are extended states, while those outside the loops (represented by black points) are localized states.
  The blue dashed curve is the ellipse given by Eq.(\ref{22}).
Standard deviations are represented by different colors of the color bar.}
\end{figure}

\begin{figure}
\begin{center}
\includegraphics[width=1.1\columnwidth]{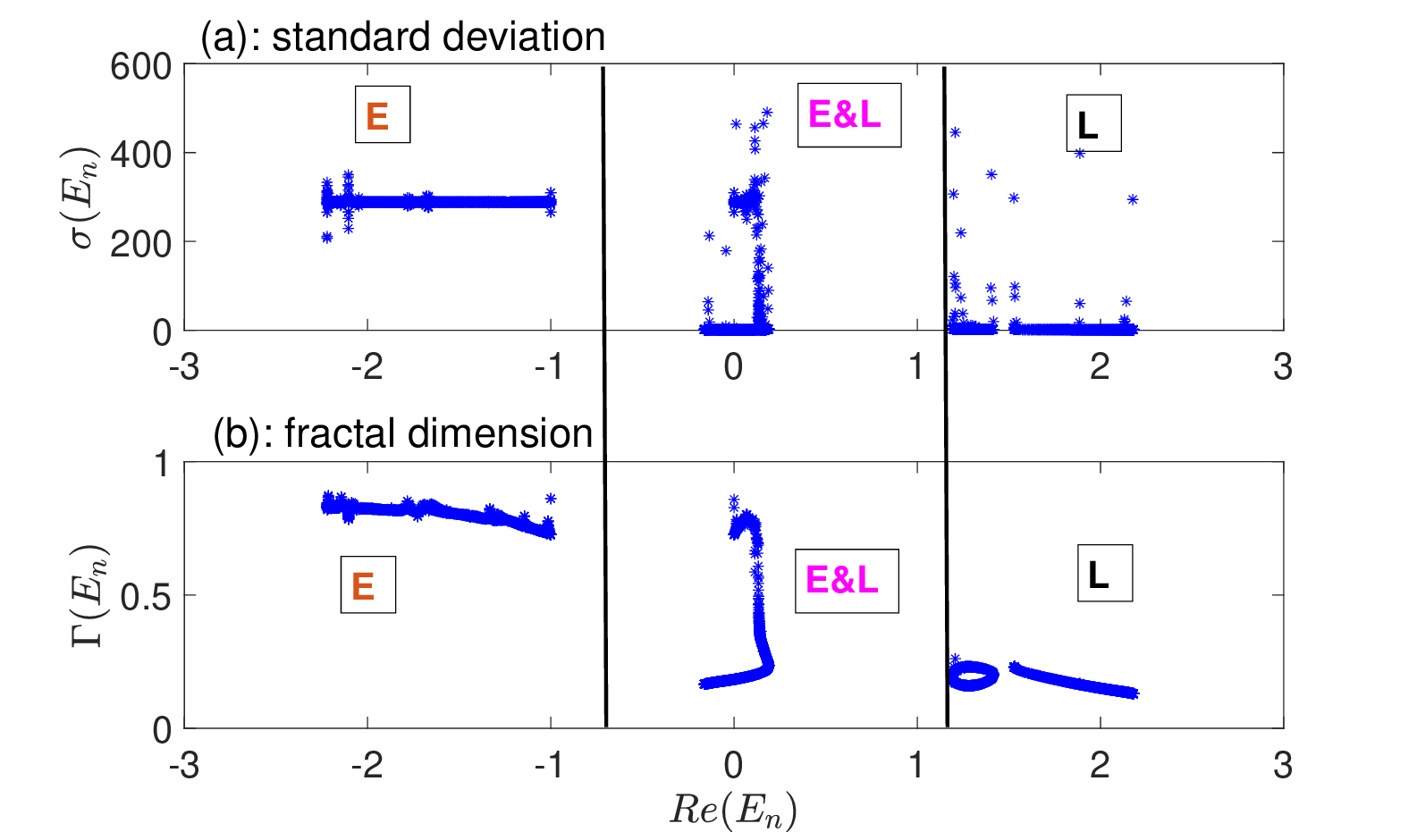}
\end{center}
\caption{(a): standard deviation and (b): fractal dimension for quasi-periodic potential of type III with $V_0=1, \theta=\pi/2, \epsilon=0.5$. Here $\textbf{L}$, $\textbf{E}$ and $\textbf{E}$ \& $\textbf{L}$ represent localized states, extended states and the mixture of localized and extended states, respectively. }
\end{figure}

In order to distinguish the localized states from the  extended states, we numerically calculate standard deviation of coordinates of eigenstates \cite{Boers2007}, i.e.,
\begin{align}\label{37}
&\sigma(n)=\sqrt{\sum_{\delta=1,2,3; j}(j-\bar{j})^2|\psi_\delta(j)|^2},
\end{align}
where $\psi_\delta$ is the nth normalized wave function and the average value of coordinate $\bar{j}$ is
\begin{align}
\bar{j}=\sum_{\delta=1,2,3; j}j|\psi_\delta(j)|^2.
\end{align}
The standard deviation $\sigma$ describes the spatial extension of the wave function in the lattice. In Fig. 4, the standard deviations  are represented with different colors of the color bar.
 The phase diagram in complex energy plane is also shown in Fig. 4. Panel (a) of Fig. 5 shows the standard deviations for all the eigenstates as a function of real parts of eigenenergy $E_n$.
 From Fig. 3, 4 and panel (a) of Fig. 5, we can see the black points in Fig. 4 are localized states which have very small standard deviations, while the yellow points are extended states which have much larger standard deviations.



In order to further investigate the properties of  the wave function,
 we also numerically calculate the fractional dimension  of the eigenstate, which is defined by
 \cite{Xiaopeng,Deng, Lizhi2024}, i.e.,
\begin{align}\label{21}
&\Gamma(n)=-\frac{\ln (\sum_{\delta j}|\psi_\delta(j)|^4)}{\ln(N)}.
\end{align}
where $\psi_\delta$ is the  nth normalized wave function.
From panel (b) of Fig. 5, we see that for localized states, the fractal dimensions are small, while for extended states, the fractal dimensions are larger.
So both the standard deviations and fractal dimensions can be used to distinguish the localized from extended states.

\section{mobility lines and mobility rings in the complex energy plane}
 From Eq. (\ref{141}), we can see that the two different expressions for the Lyapunov exponent are separated  by the equation $|\epsilon| = \ln \left| \frac{1 + \sqrt{1 - \alpha^2}}{\alpha} \right|$. This equation forms an ellipse in the complex energy plane (please refer to the detailed proof in the Appendix). The equation of the ellipse can be explicitly given by:
 \begin{align}\label{22} \frac{(x \cos \theta + y \sin \theta)^2}{\left( \frac{V_0}{2} \cosh \epsilon \right)^2} + \frac{(y \cos \theta - x \sin \theta)^2}{\left( \frac{V_0}{2} \sinh \epsilon \right)^2} = 1, \end{align}
  where $x = Re(E)$ and $y = Im(E)$. We see that the angle between the major axis of ellipse and positive real axis of $E$ is $\theta$  (as shown by the blue dashed curves in Figs. 4 and 6).
  When $\theta=0$ and $\epsilon=0$, this ellipse is reduced to a line segment, given by $-V_0/2<E<V_0/2$. If $E$ lies on the line segment,  the parameter $\alpha$ would satisfies $|\alpha|=|\frac{V_0}{2E}|>1$, which defines the unbounded Hermitian cases \cite{zhangyicai2024}.
   While for  general non-Hermitian cases, this ellipse divides the entire complex energy plane into two parts.

\subsection{mobility lines inside the ellipse $\frac{(x \cos\theta+y\sin\theta)^2}{(\frac{V_0}{2}\cosh\epsilon)^2}+\frac{(y \cos\theta-x \sin\theta)^2}{(\frac{V_0}{2}\sinh\epsilon)^2}<1$}
By Eq.(\ref{141}), we find that inside the ellipse, the Lyapunov exponent is given by
\begin{align}
\gamma(E,\epsilon)=\frac{1}{2}\textrm{Max}\{0,\ln(\frac{|P\pm\sqrt{P^2-4\alpha^2}|}{2|\alpha|})\}.
\end{align}
Then the mobility edges, denoted by $E_c$ which separates the localized states from the extended states, can be determined  as follows:
\begin{align}\label{241}
\frac{1}{2}\textrm{Max}\{0,\ln(\frac{|P\pm\sqrt{P^2-4\alpha^2}|}{2|\alpha|})\}=0.
\end{align}
By Eq.(\ref{241}), then we find
\begin{align}\label{24}
E_c\in[-2,-1] \cup [0, 1].
\end{align}
We observe that Eq.(\ref{24}) is independent of the phase $\theta$ and $i\epsilon$. To be precise, here the mobility edges are determined by the intersection of two sets: $\{E=x+yi | \frac{(x \cos\theta+y\sin\theta)^2}{(\frac{V_0}{2}\cosh\epsilon)^2}+\frac{(y \cos\theta-x \sin\theta)^2}{(\frac{V_0}{2}\sinh\epsilon)^2}<1\}$ and $\{E | E \in[-2,-1] \cup [0, 1]\}$, which are represented by the green solid line segments inside the blue dashed ellipses in Figs. 4 and 6.

Furthermore, it has been observed that within the ellipse, if imaginary part of $E_n$ is not zero [$Im(E_n)\neq0$], the Lyaponov exponent is always positive, indicating localized states (represented by the black points inside the blue dashed ellipses in Figs. 4 and 6). Only when $E_n=E_c$, these states become extended (represented by the yellow points on the green line segments in Figs. 4 and 6).
This implies that only when the eigenenergies lie on the mobility lines, their corresponding eigenstates are extended states.

By expanding the Lyapunov exponent [Eq.(\ref{141})] near the mobility edges $E_c$ (the points on the green line segments in Figs. 4 and 6), we can determine the behavior of Lyapunov, i.e.,
\begin{align}
\gamma(E,\epsilon)\propto |E-E_c|^\nu\rightarrow0, \ as \ E \rightarrow E_c.
\end{align}
Then the localization length is
\begin{align}
\xi(E)\equiv1/\gamma(E,\epsilon)\propto |E-E_c|^{-\nu}\rightarrow\infty,\ as \ E \rightarrow E_c,
\end{align}
where $\nu$ is the  critical index of localized length.
We find that the critical index $\nu$ depends on the position of  $E_c$.
For example,  if $E_c=0$, when $E$ approaches $E_c=0$ , the critical index  $\nu=1/2$, i.e.,
\begin{align}
\xi(E)\propto |E-E_c|^{-1/2}\rightarrow0, \ as \ E\rightarrow E_c=0.
\end{align}
While when $E$ approaches  $E_c=1/3$, the critical index  $\nu=1$, i.e.,
\begin{align}
\xi(E)\propto|E-E_c|^{-1}\rightarrow0, \ as \ E\rightarrow E_c=1/3.
\end{align}
This shows that, differently from Hermitian cases \cite{yicai},  here the critical index $\nu$ is not a constant, but rather varies depending on the  positions of mobility edges in the complex energy plane.

\subsection{mobility rings outside the ellipse $\frac{(x \cos\theta+y\sin\theta)^2}{(\frac{V_0}{2}\cosh\epsilon)^2}+\frac{(y \cos\theta-x \sin\theta)^2}{(\frac{V_0}{2}\sinh\epsilon)^2}>1$}
When the energy is outside the ellipse, by Eq.(\ref{141}), the Lyapunov exponent is given by
\begin{align}
\gamma(E,\epsilon)=\frac{1}{2}\textrm{Max}\{0,|\epsilon|+\ln(\frac{|P\pm\sqrt{P^2-4\alpha^2}|}{2|1+\sqrt{1-\alpha^2}|})\}.
\end{align}
Then the mobility edges, denoted by $E_c$, can be determined  as follows:
\begin{align}\label{30}
\frac{1}{2}\textrm{Max}\{0,|\epsilon|+\ln(\frac{|P\pm\sqrt{P^2-4\alpha^2}|}{2|1+\sqrt{1-\alpha^2}|})\}=0.
\end{align}
It is found that the mobility edges form closed loops in complex energy plane, named as mobility rings \cite{Lizhi2024} which are represented by the green solid curves in Fig. 4 and 6.


When the eigenvalues are outside the mobility rings,   their Lyapunov exponents  are always positive, then these states  corresponds localized states (see black points outside the blue ellipses in Figs. 4 and 6).
Only when eigenvalues are inside the mobility rings, their Lyapunov exponents are zero, and the states correspond extended states (see yellow points outside the blue ellipses in Figs. 4 and 6).

\begin{figure}
\begin{center}
\includegraphics[width=1.1\columnwidth]{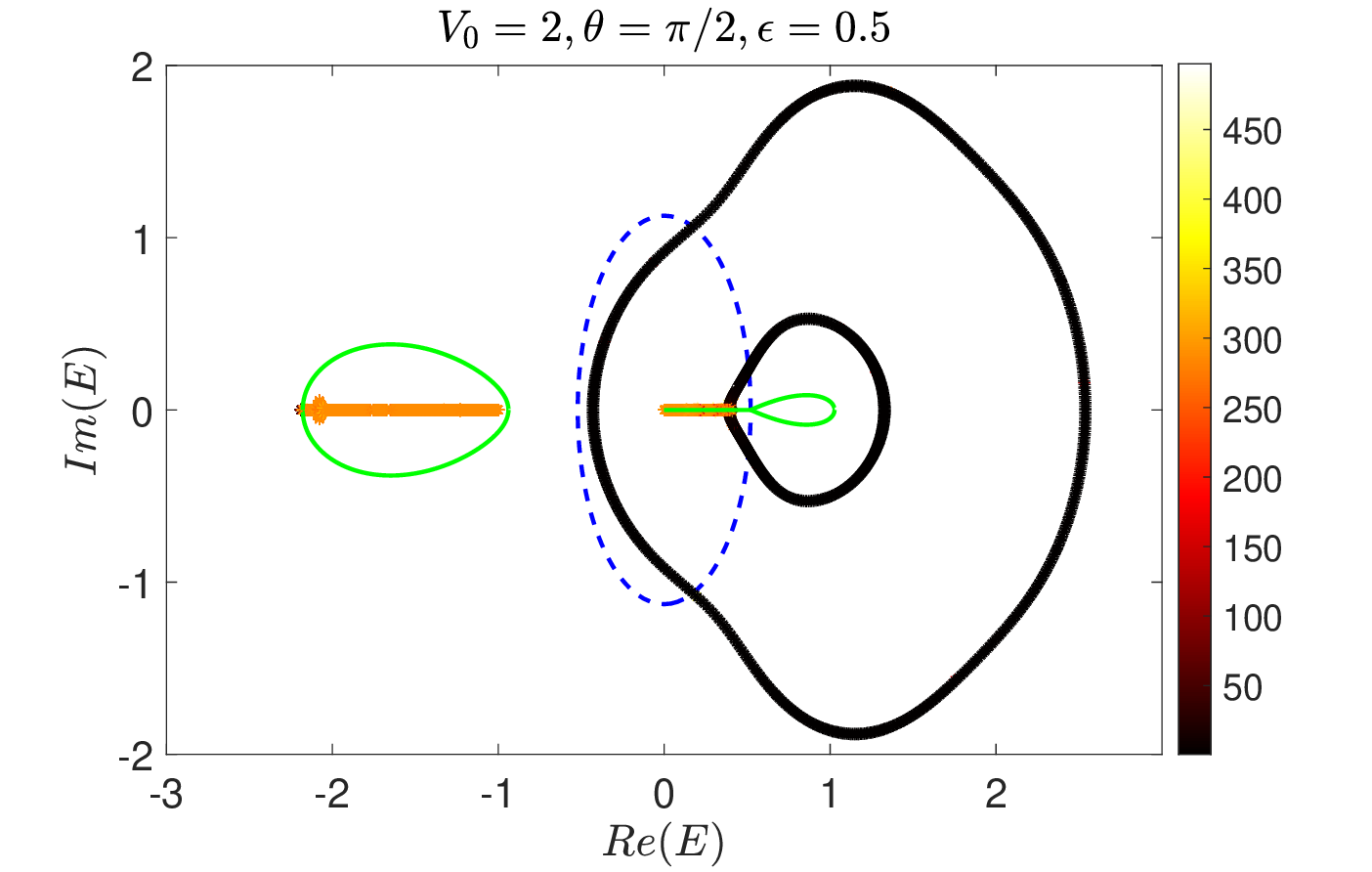}
\end{center}
\caption{The phase diagram for quasi-periodic potential of type III in the complex $E$ plane for $V_0=2$, $\theta=\pi/2$, $\epsilon=0.5$ and lattice size $3N=3\times 1000=3000$.
 The phase boundaries (mobility edges $E_c$) are indicated by the green solid line segment and curves, and their equations are given by Eqs. (\ref{24}) and (\ref{30}).
 These states which are within the green loops and on the green line segment (yellow color points)  are extended states, while those  outside the loops are localized states (black color points).
  The blue dashed curve is the ellipse given by Eq.(\ref{22}).}
\end{figure}

\subsection{mobility rings in  quasi-periodic potential of Type II}

In this subsection, for the sake of completeness, we investigate the Anderson localization problem in the flat band lattice model for a  non-Hermitian quasi-periodic potential of Type II. We assume that the potential energy, denoted as $V_p$, takes the following form in the spin basis $|1,2,3\rangle$:
\begin{align}\label{A1}
V_p=\sum_n V_{22}(n)b^{\dag}_{n}b_{n}.
\end{align}
Note that the potential only appears in the basis element $|2\rangle$ (or sublattice B). Throughout the manuscript, we will refer to this type of potential as ``type II" potential \cite{24}.

Furthermore, like $V_{11}(n)$, we assume  $V_{22}(n)$ is  also a  non-Hermitian quasi-periodical potential, i.e.,
\begin{align}\label{A2}
V_{22}(n)=V_0e^{i\theta}\cos(2\pi\beta n+\phi+i\epsilon).
\end{align}

The Schr\"{o}dinger equation can be written as
\begin{align}\label{A3}
&-it[\psi_{2}(n+1)-\psi_{2}(n-1)]/\sqrt{2}=[E-m]\psi_{1}(n),\notag\\
&-it[\psi_{1}(n+1-\psi_{1}(n-1)+\psi_{3}(n+1)-\psi_{3}(n-1)]/\sqrt{2}\notag\\
&=[E-V_{22}(n)]\psi_{2}(n),\notag\\
&-it[\psi_{2}(n+1)-\psi_{2}(n-1)]/\sqrt{2}=[E+m]\psi_{3}(n).
\end{align}
Eliminating wave functions for first and third components, we get an effective Schr\"{o}dinger  equation for $\psi_2$
\begin{align}\label{A4}
&-t^2[\psi_{2}(n+2)-2\psi_{2}(n)+\psi_{2}(n-2)]\notag\\
&+\frac{(E^2-m^2)V_{22}(n)}{E}\psi_{2}(n)=[E^2-m^2]\psi_{2}(n).
\end{align}
Further introducing an  effective hopping $\tilde{t}$, an effective energy $\tilde{E}$ and an effective potential strength $\lambda$
\begin{align}\label{A5}
&\tilde{t}\equiv t^2,\notag\\
&\tilde{E}\equiv -E^2+m^2+2t^2,\notag\\
&\lambda\equiv V_0e^{i\theta}(-E^2+m^2)/(2E),
\end{align}
Eq.(\ref{A4}) becomes a non-Hermitian Aubry-Andr\'{e} (AA) model, i.e.,
\begin{align}\label{A6}
&\tilde{t}[\psi_{2}(n+2)+\psi_{2}(n-2)]+2\lambda \cos(2\pi\beta n+\phi+i\epsilon)\psi_{2}(n)\notag\\
&=\tilde{E}\psi_{2}(n).
\end{align}

%
%
%
%
%
%

Using Avila's theory,   the Lyapunov exponent for the AA model  is given by (see the Appendix)
\begin{align}\label{A11}
&\gamma(E,\epsilon)=\frac{1}{2}\textrm{Max}\{0,|\epsilon|+\ln(|\frac{\lambda}{\tilde{t}}|)\}\notag\\
&=\frac{1}{2}\textrm{Max}\{0,|\epsilon|+\ln(|\frac{V_{0}(E^2-m^2)}{2Et^2}|)\}.
\end{align}

\begin{figure}
\begin{center}
\includegraphics[width=1.1\columnwidth]{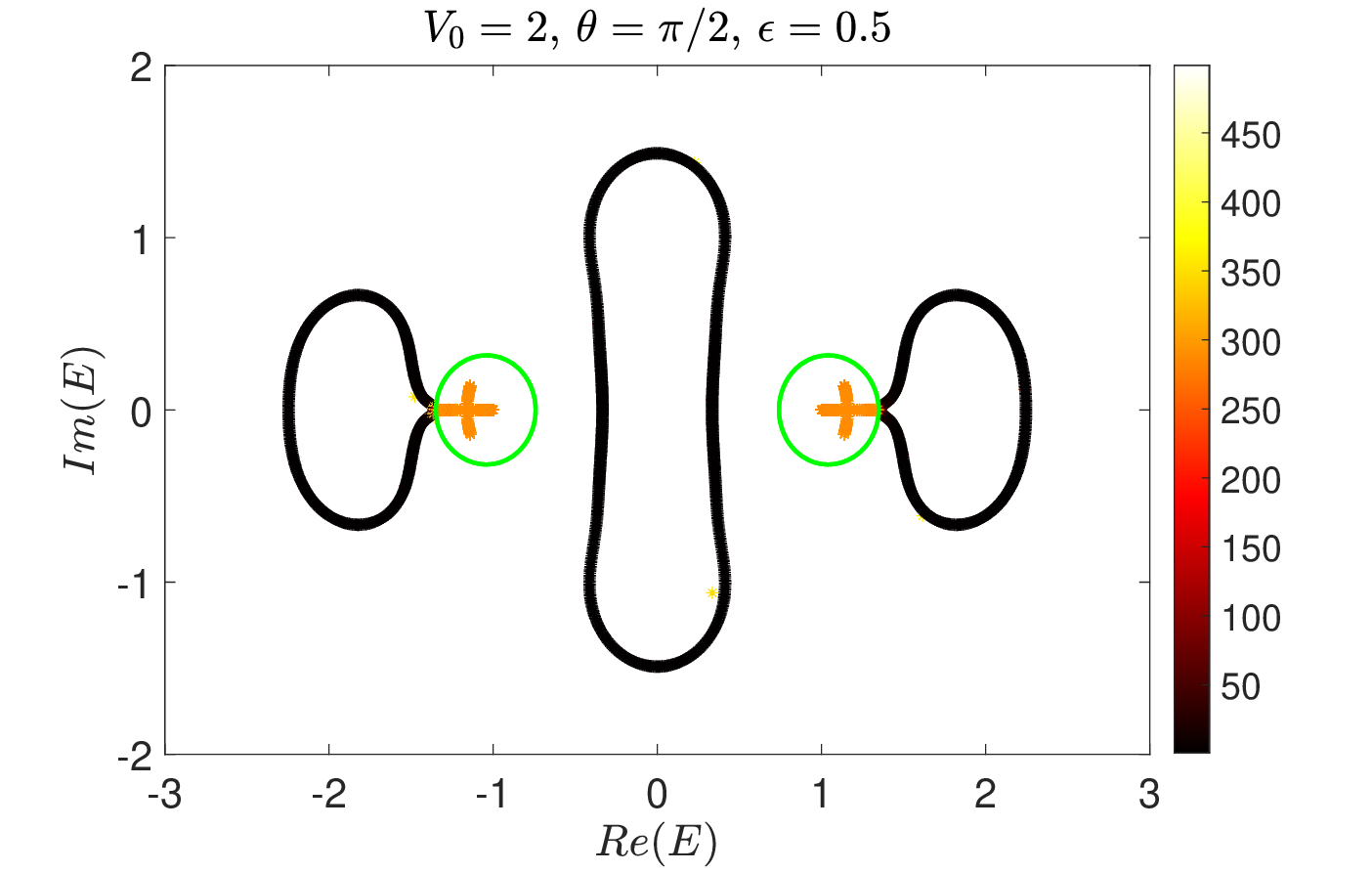}
\end{center}
\caption{Phase diagram for quasi-periodic potential of type II with $V_0=2,\theta=\pi/2,\epsilon=0.5$. The phase boundaries (mobility edges $E_c$) are indicated by the green solid curves, and their equations are given by Eq. (\ref{A12}).  The states which are inside the green rings (yellow color points) are extended states, while those  outside the rings are localized states (black color points).
The standard deviations are represented by different colors of the color bar.}
\end{figure}

\begin{figure}
\begin{center}
\includegraphics[width=1.1\columnwidth]{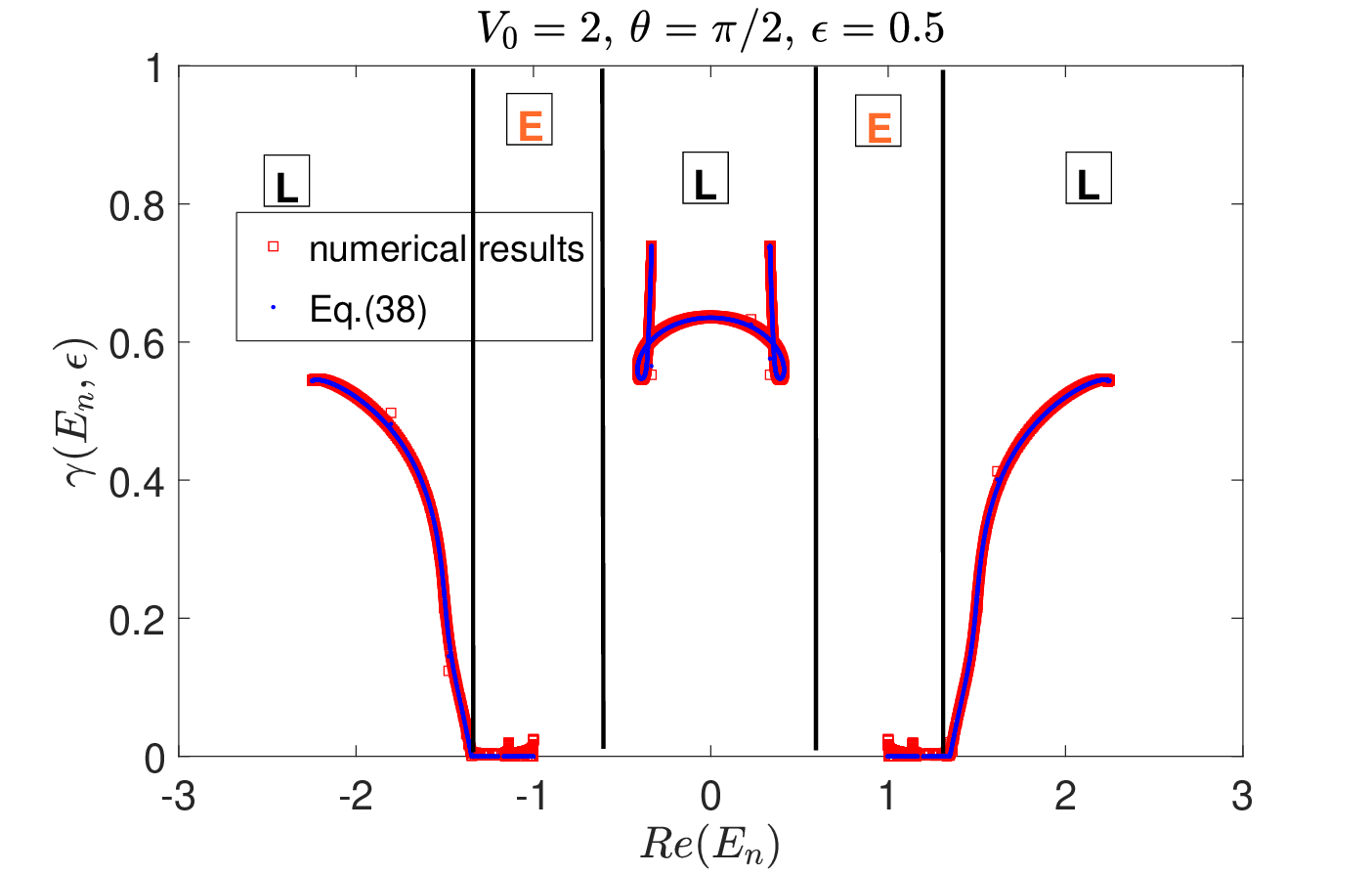}
\end{center}
\caption{Lyapunov exponent for quasi-periodic potential of type II with  $V_0=2, \theta=\pi/2, \epsilon=0.5$. The red squares are obtained through numerical calculation, while the blue dots are obtained using Eq.(\ref{A11}). It is evident that the two sets of discrete points overlap perfectly. Here $\textbf{L}$ and  $\textbf{E}$ represent localized states, extended states, respectively. }
\end{figure}

\begin{figure}
\begin{center}
\includegraphics[width=1.1\columnwidth]{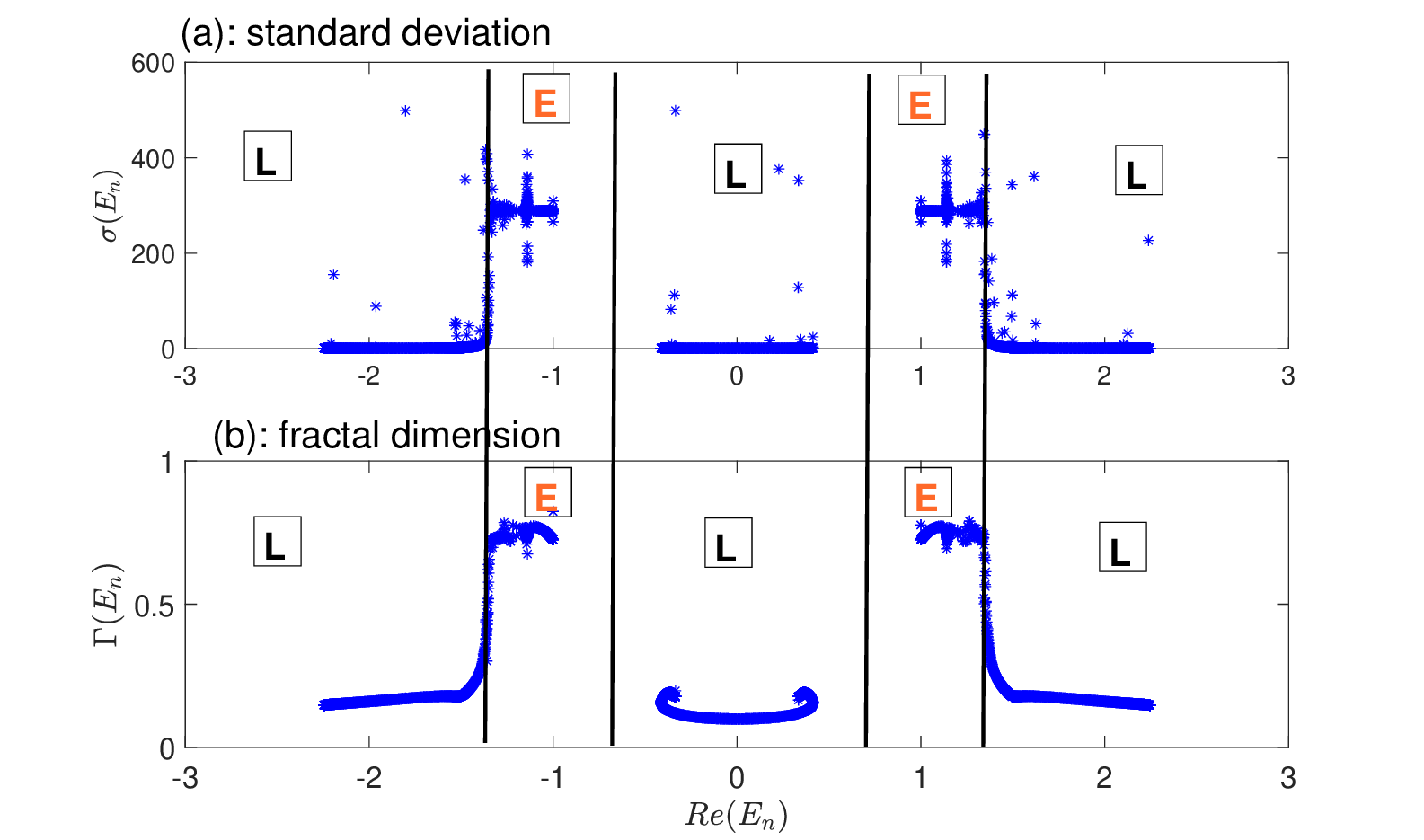}
\end{center}
\caption{(a): standard deviation and (b):fractal dimension for quasi-periodic potential of type II  with $V_0=2, \theta=\pi/2, \epsilon=0.5$.  Here $\textbf{L}$ and  $\textbf{E}$ represent localized states, extended states, respectively.}
\end{figure}

By Eq.(\ref{A11}), the mobility edges are determined by
\begin{align}\label{A12}
&\gamma(E_c)=\frac{1}{2}\textrm{Max}\{0,|\epsilon|+\ln(|\frac{\lambda}{\tilde{t}}|)\}=0,\notag\\
&\Rightarrow |\epsilon|+\ln(|\frac{\lambda}{\tilde{t}}|)=0,\notag\\
&\Rightarrow(x-\frac{x}{x^2+y^2})^2+(y+\frac{y}{x^2+y^2})^2=\frac{4t^4e^{-2|\epsilon|}}{m^2V_{0}^{2}},
\end{align}
where $x=Re(E_c)/m$ and $y=Im(E_c)/m$.
The mobility edges form mobility rings in the complex energy plane, represented by the green solid curves in Fig. 7.
The phase diagram is also presented in Fig. 7, and the different colors of color bar represent different values of the standard deviations.

 According to Eq. (\ref{A12}), the mobility rings are not affected by the parameter $\theta$.  As the absolute value of $\epsilon$ increases, the sizes of the mobility rings decrease. In particular, when $|\epsilon|\rightarrow \infty$, the mobility rings shrink into two points and  disappear eventually.

 Using transfer matrix method, we  also performed numerical calculations for the Lyapunov exponents (represented by the set of red squares in Fig. 8). The blue dots in Fig. 8 represent the values obtained from Eq. (\ref{A11}) for all eigenenergies. The perfect overlap of the two sets of discrete points indicates a strong agreement between the analytical and numerical results.

We also calculated the standard deviations and  fractal dimensions for all the eigenstates. The results are shown in Fig. 9. From Fig. 9, it is evident  that the standard deviations and fractal dimensions are significantly smaller for localized states compared to extended states.

\section{summary}

In conclusion, we investigate the localization properties of a one-dimensional flat band lattice model with a non-Hermitian quasiperiodic on-site potential.
Our findings indicate that the presence of a non-Hermitian potential leads to the disappearance of critical states and critical regions. By utilizing Avila's theory, we determine the exact Lyapunov exponents and mobility edges. Our results show that the mobility edges not only form mobility rings, but also  mobility lines in the complex energy plane. Within the mobility rings, the states are extended, while the states outside the rings are localized. For mobility line cases, only when the eigenenergies lie on these lines, their corresponding eigenstates are extended states.
 Furthermore, our research suggests that the critical index of the localization length  can vary depending on the positions of mobility edges in the complex energy plane.

\section*{Acknowledgements}
This work was supported by the NSFC under Grants Nos.
11874127, 12074180,  12174077, Guangdong Basic and Applied Basic Research Foundation under No. 2023A1515010698, the Joint Fund with
Guangzhou Municipality under No.202201020137, and the Starting Research Fund from
Guangzhou University under Grant No.
RQ 2020083.


\section*{Appendix}

\subsection{ derivations of Eq.(\ref{141}) and Eq.(\ref{A11})}
In this subsection of Appendix, we use the transfer matrix method \cite{Sorets1991,Davids1995} to calculate the Lyapunov exponent.
First of all,   let us assume that the system [Eq.(\ref{11})] is a half-infinite lattice with left-hand end sites at $n=0$ and $n=2$.
Further using Eq.(\ref{11}), starting with $\psi(0)$ and $\psi(2)$ of left-hand end sites, we can  calculate the wave function for the entire system \cite{yicai}, i.e.,
\begin{align}
\Psi(2L)=T(2L)T(2L-2)...T(4)T(2)\Psi(0)
\end{align}
where transfer matrix
\begin{align}\label{A45}
T(n)\equiv\left[\begin{array}{ccc}
\frac{\tilde{E}}{\tilde{t}}-\frac{2\lambda}{\tilde{t}} \frac{\cos(2\pi\beta n+\phi+i\epsilon)}{1-\alpha \cos(2\pi\beta n+\phi+i\epsilon) } &-1  \\
1&0\\
  \end{array}\right],
\end{align}
and
\begin{align}
\Psi(n)\equiv\left[\begin{array}{ccc}
\psi(n+2)  \\
\psi(n)\\
  \end{array}\right].
\end{align}

The Lyapunov exponent is defined by
\begin{align}\label{V0}
&\gamma(E,\epsilon)=\lim_{L \rightarrow \infty }\frac{\ln(|\Psi(2L)|/|\Psi(0)|)}{2L}\notag\\
&=\lim_{L\rightarrow \infty}\frac{\ln(|T(2L)T(2L-2)...T(4)T(2)\Psi(0)|/|\Psi(0)|)}{2L}
\end{align}
where $L$ is a positive integer,
and the modulus is defined as
\begin{align}
|\Psi(n)|=\sqrt{|\psi(n+2)|^2+|\psi(n)|^2}.
\end{align}

The transfer matrix Eq.(\ref{A45}) can be further
written as a product of two parts, i.e., $T_n= A_nB_n$, where
   \begin{align}
&A_n=\frac{1}{1-\alpha \cos(2\pi\beta n+\phi+i\epsilon)},\notag\\
&B_n=\left[\begin{array}{ccc}
B_{11} &B_{12} \\
B_{21}&0\\
  \end{array}\right],
\end{align}
with $B_{11}=\frac{\tilde{E}}{\tilde{t}}[1-\alpha \cos(2\pi\beta n+\phi+i\epsilon)]-2\lambda \cos(2\pi\beta n+\phi+i\epsilon)/\tilde{t}$, and $B_{21}=-B_{12}=1-\alpha \cos(2\pi\beta n+\phi+i\epsilon)$..

Now  the Lyapunov exponent can be written as
\begin{align}
\gamma(E,\epsilon)=\gamma_A(E,\epsilon)+\gamma_B(E,\epsilon),
\end{align}
where
\begin{align}
&\gamma_A(E,\epsilon)=\lim_{L\rightarrow \infty}\frac{\ln(|A(2L)A(2L-2)...A(4)A(2)|)}{2L},
\end{align}
and $\gamma_B(E)$ are given by
\begin{align}
&\gamma_B(E,\epsilon)\notag\\
&=\lim_{L\rightarrow \infty}\frac{\ln(|B(2L)B(2L-2)...B(4)B(2)\Psi(0)|/|\Psi(0)|)}{2L}.
\end{align}

In the following, we will utilize Avila's theory \cite{Avila2015} to obtain the Lyapunov exponent.
 First of all, the ergodicity of the map $\phi\longrightarrow 2\pi\beta n+\phi$ allows us write  $\gamma_A(E)$ as an integral over phase $\phi$ \cite{Longhi2019,Liu2021, YONGJIAN1}, consequently
\begin{align}
&\gamma_A(E,\epsilon)=\frac{1}{2\times2\pi}\int_{0}^{2\pi} d\phi \ln(|\frac{1}{1-\alpha \cos(\phi+i\epsilon)}|)\notag\\
\notag \\
&=
\left\{\begin{array}{cccc}
&&-\frac{1}{2}\ln(\frac{|1+\sqrt{1-\alpha^2}|}{2}), \\
\\
 &&  for \ \ \ \ |\epsilon|<
\ln(|\frac{1+\sqrt{1-\alpha^2}}{\alpha}|)\\
\\
&&-\frac{|\epsilon|}{2}-\frac{1}{2}\ln(\frac{|\alpha|}{2}),  \\
\\
&& for  \ \ \ \ |\epsilon| \geq \ln(|\frac{1+\sqrt{1-\alpha^2}}{\alpha}|) \end{array}\right.
\end{align}
Note: in contrast to the Hermitian case \cite{yicai,zhangyicai2024}, here $\alpha$ is a complex number in general [see Eq. (\ref{10})]. In the calculation of the above integral, we utilize Jensen's formula for a complex variable function, which is related to the zeros, the poles and the value of the origin of the complex variable function \cite{Jensen,Lingrui}.

Next we take $|\epsilon|\rightarrow +\infty$, then
 \begin{align}
B_n=\frac{e^{-isgn(\epsilon)(2\pi\beta n+\phi)+|\epsilon|}}{2}\left[\begin{array}{ccc}
\frac{-(\alpha \tilde{E}+2\lambda)}{\tilde{t}} &\alpha \\
-\alpha&0\\
  \end{array}\right]+O(1),
\end{align}
where $sgn(\epsilon)$ is the sign function of $\epsilon$.
Then for large $|\epsilon|$, i.e.,  $|\epsilon|\gg1$, $\gamma_B(E,\epsilon)$ is determined by the largest eigenvalue (in absolute value) of $B_n$, i.e.,
\begin{align}
&\gamma_B(E,\epsilon)=
\frac{|\epsilon|}{2}+\frac{1}{2}\textrm{Max}\{\ln(\frac{|P\pm \sqrt{P^2-4\alpha^2}|}{4})\},
\end{align}
where
\begin{align}\label{56}
P=\frac{\alpha \tilde{E}+2\lambda}{\tilde{t}}.
\end{align}

Using the facts that $\gamma(E,\epsilon)\geq0$  and $\gamma(E,\epsilon)$ is  piecewise linear function of $\epsilon$ \cite{Avila2015},  one can get
\begin{align}\label{A57}
\gamma(E,\epsilon)=\textrm{Max}\{0,\gamma_A(E,\epsilon)+\gamma_B(E,\epsilon)\},\notag\\
  \notag \\
=\left\{\begin{array}{cccc}
&&\frac{1}{2}\textrm{Max}\{0,|\epsilon|+\ln(\frac{|P\pm\sqrt{P^2-4\alpha^2}|}{2|1+\sqrt{1-\alpha^2}|})\}, \\
\\
 &&  for \ \ \ \ |\epsilon|<
\ln(|\frac{1+\sqrt{1-\alpha^2}}{\alpha}|)\\
\\
&&\frac{1}{2}\textrm{Max}\{0,\ln(\frac{|P\pm\sqrt{P^2-4\alpha^2}|}{2|\alpha|})\},  \\
\\
&& for  \ \ \ \ |\epsilon| \geq \ln(|\frac{1+\sqrt{1-\alpha^2}}{\alpha}|)\\
  \end{array}\right.
\end{align}
which is  Eq. (\ref{141}) in main text.

 When $\alpha$ approaches 0, i.e. $\alpha\rightarrow0$, Eq. (\ref{11}) becomes formally equivalent to Eq. (\ref{A6}). According to Eq. (\ref{56}), $P$ approaches $2\lambda/\tilde{t}$, and for any arbitrary $\epsilon$, the inequality $|\epsilon|< \ln(|\frac{1+\sqrt{1-\alpha^2}}{\alpha}|)\rightarrow +\infty$ always holds. Therefore,  Eq. (\ref{A57}) can be simplified to Eq. (\ref{A11}) in the main text. This means that we can view Eq. (\ref{A11}) as a special case of Eq. (\ref{A57}).

  It is important to note that although the two definitions of effective $\lambda$ in Eq. (\ref{10}) and Eq. (\ref{A5}) are different, the derivation of the Lyaponov exponent only depends on the function form of the quasiperiodic potential, rather than the specific value of $\lambda$. Therefore, the arguments presented above are still valid.

\subsection{ The proof of  $|\epsilon|=
\ln(|\frac{1+\sqrt{1-\alpha^2}}{\alpha}|)$ is an ellipse}
From the definition of $\alpha$ Eq. (\ref{10}), let us set
\begin{align}\label{A58}
\alpha=\frac{V_0e^{i\theta}}{2E}\equiv\frac{1}{z}\Rightarrow z=\frac{2e^{-i\theta}E}{V_0},
\end{align}
and
then the equation becomes
\begin{align}
|\epsilon|=
\ln(|\frac{1+\sqrt{1-\alpha^2}}{\alpha}|)\Rightarrow |z+\sqrt{z^2-1}|=e^{|\epsilon|}.
\end{align}

Next we represent complex number $z$ as $z=x+yi$ where $x$ and $y$ are real and imaginary parts of $z$, respectively.
Then, the above equation is
\begin{align}
|z+\sqrt{z^2-1}|=|x+yi+\sqrt{x^2-y^2-1+2xyi}|=e^{|\epsilon|}.
\end{align}
Further we write $x^2-y^2-1+2xyi$ as a square of a complex number $m+ni$, i.e.,
\begin{align}\label{A61}
x^2-y^2-1+2xyi=(m+ni)^2,
\end{align}
where $m, n$ are two real number, then we get
\begin{align}
&|z+\sqrt{z^2-1}|=|x+yi+\sqrt{x^2-y^2-1+2xyi}|\notag\\
&=|x+yi+m+ni|=e^{|\epsilon|}.
\end{align}
So we get
\begin{align}\label{A63}
(x+m)^2+(y+n)^2=e^{2|\epsilon|}.
\end{align}
  By Eq. (\ref{A61}), we have
\begin{align}\label{A631}
&x^2-y^2-1=m^2-n^2,\notag\\
&xy=mn.
\end{align}

Solving Eq. (\ref{A631}) , we get
\begin{align}\label{A641}
&m=\frac{\sqrt{-1+x^2-y^2+\sqrt{1-2x^2+x^4+2y^2+2x^2y^2+y^4}}}{\sqrt{2}},\notag\\
&n=\frac{xy}{m}.
\end{align}

Further we assume $m=kx$ and $n=y/k$ where $k$ is a real number to be determined.   We find that  Eq. (\ref{A641}) becomes an equation of ellipse:
\begin{align}\label{A651}
(1-k^2)x^2+(\frac{1}{k^2}-1)y^2=1.
\end{align}
and Eq. (\ref{A63}) is also an equation of ellipse, i.e,
\begin{align}\label{A66}
(1+k)^2x^2+(1+1/k)^2y^2=e^{2|\epsilon|}.
\end{align}

These above two equations should lead to a same ellipse, then comparing Eq. (\ref{A651}) with Eq. (\ref{A66}), we find
\begin{align}
(1+k)^2e^{-2|\epsilon|}=1-k^2\Rightarrow k=\tanh(|\epsilon|),
\end{align}
then
\begin{align}
&1-k^2=\frac{1}{\cosh^2(\epsilon)},\notag\\
&\frac{1}{k^2}-1=\frac{1}{\sinh^2(\epsilon)}.
\end{align}

So the ellipse Eq. (\ref{A651}) is
\begin{align}\label{A70}
\frac{x^2}{\cosh^2(\epsilon)}+\frac{y^2}{\sinh^2(\epsilon)}=1.
\end{align}
From Eq. (\ref{A58}), we have
\begin{align}
&z=\frac{2e^{-i\theta}E}{V_0}\Rightarrow x+yi\notag\\
&=\frac{2}{V_0}\{Re(E) \cos\theta+Im(E) \sin\theta\notag\\
&+i[-Re(E) \sin\theta+Im(E) \cos\theta]\},
\end{align}
where complex energy  $E=Re(E)+iIm(E)$.
Then
\begin{align}\label{72}
&x=\frac{2}{V_0}[Re(E) \cos\theta+Im(E) \sin\theta],\notag\\
&y=\frac{2}{V_0}[(-Re(E) \sin\theta+Im(E) \cos\theta)].
\end{align}
Substituting Eq. (\ref{72}) into  Eq. (\ref{A70}), the equation of ellipse becomes
\begin{align}\label{73}
&\frac{[Re(E) \cos\theta+Im(E) \sin\theta]^2}{(\frac{V_0}{2}\cosh\epsilon)^2}+\frac{[Im(E) \cos\theta-Re(E) \sin\theta]^2}{(\frac{V_0}{2}\sinh\epsilon)^2}=1.
\end{align}

Finally if we reset $Re(E)\equiv x$ and $Im(E)\equiv y$ in Eq. (\ref{73}), then we get the equation of the ellipse
\begin{align}
&\frac{(x \cos\theta+y\sin\theta)^2}{(\frac{V_0}{2}\cosh\epsilon)^2}+\frac{(y \cos\theta-x \sin\theta)^2}{(\frac{V_0}{2}\sinh\epsilon)^2}=1,
\end{align}
where $E=x+yi$.

\end{document}